# Time-Variation of Newton's Constant and the Age of Globular Clusters

**S. Degl'Innocenti**[1,2], **G. Fiorentini**[2,3], **G.G. Raffelt**[4], **B. Ricci**[5,2], **and A. Weiss**[1]

[1] Max-Planck-Institut für Astrophysik, Karl-Schwarzschild-Str. 1, D-85740 Garching, Germany
[2] Istituto Nazionale di Fisica Nucleare, Sezione di Ferrara, I-44100 Ferrara, Italy
[3] Dipartimento di Fisica dell'Università di Ferrara, I-44100 Ferrara, Italy
[4] Max-Planck-Institut für Physik, Föhringer Ring 6, D-80805 München, Germany
[5] Scuola di Dottorato, Università di Padova, Italy

**Abstract.** A variation of Newton's constant $G$ over cosmological time scales would modify the main-squence time of globular cluster (GC) stars. We have calculated the evolution of low-mass stars typical for GCs both for standard nonvarying $G$ and under the assumption of a linear variation of $G$. The age of the isochrones resulting from the latter models then was chosen such that the isochrones mimicked the standard ones at the turnoff. Assuming that the true age of GCs is between 8 and 20 Gyr, and because their apparent age is between 14 and 18 Gyr, we find that today $-35 \times 10^{-12}\,\mathrm{yr}^{-1} \lesssim \dot{G}/G \lesssim 7 \times 10^{-12}\,\mathrm{yr}^{-1}$. The upper limit (gravity weaker in the past) is competitive with direct present-day bounds from celestial mechanics. Within independently determined $\dot{G}/G$ limits a time-varying $G$ as an explanation for the discrepancy between the cosmic expansion age and the apparent GC ages is conceivable.

## 1. Introduction

Some constants of nature are thought to be more constant than others. For example, the "cosmological constant" $\Lambda$ is now routinely interpreted as representing the vacuum energy of quantum fields. In the framework of inflationary cosmologies, $\Lambda$ is assumed to be a dynamical variable which initially drives a de Sitter expansion of the universe, and which later evolves to its present-day small or vanishing value. Another example are the masses of elementary particles which are thought to arise from the interaction with a cosmological background field, the vacuum expectation value $\langle\Phi\rangle$ of the Higgs field $\Phi$. A nonvanishing value for $\langle\Phi\rangle$ appears only when the universe has cooled to $T \lesssim 250\,\mathrm{GeV}$; at earlier epochs the particles have vanishing vacuum masses.

Therefore, it may not be too absurd to imagine the possibility that another dimensionful constant of nature, Newton's constant $G$, could also represent a dynamical degree of freedom and thus vary on cosmological time scales. Indeed, certain extensions of general relativity, notably the Brans-Dicke scalar-tensor theory, predict such a time variation. Independently of specific models it is worthwhile to derive limits on, or find evidence for a possible time variation of the strength of the gravitational force. Three methods have been discussed in the literature to constrain or discover a putative $G$ time variation: Direct constraints on a present-day $\dot{G}$ from the orbits of celestial bodies, a constraint on the value of $G$ at the time of big bang nucleosynthesis (BBN), and stellar-evolution limits, notably from the properties of the Sun. We will review previous results obtained by these methods in Sect. 2.

Teller (1948) was the first to stress that a decrease (increase) of $G$ would have caused stars to burn faster (slower) in the past than standard. For the Sun this would mean that it is effectively more (less) evolved than in the standard picture. Recently it has been shown that data on solar p- and g-mode frequencies are the most powerful tool to constrain this possibility (Demarque et al. 1994; Guenther et al. 1995). However, a potentially more sensitive method to test a modified speed of stellar evolution in the past is provided by globular clusters (GCs) which have probed $G$ since much earlier times than the Sun. Apparently, the only detailed investigation of this case was conducted in an unpublished Ph.D. thesis about twenty years ago (Prather 1976). It is the purpose of our present note to perform a new study of what can be learned about $\dot{G}$ from a comparison between the apparent ages of GCs and the accepted range of possible true ages.

In Sect. 2 we begin with a review of previous $\dot{G}/G$ limits. In Sect. 3 we discuss GC color-magnitude diagrams and evolutionary time scales in the presence of a $G$ time variation. We derive a new limit on $\dot{G}/G$ from GC ages. In Sect. 4 we discuss homology relations between $G$ and the main-sequence luminosity.

## 2. Previous limits on time-varying gravity

### 2.1. Celestial mechanics

Particularly precise data on the orbits of celestial bodies exist in the solar system from laser ranging of the moon and radar ranging of planets, notably by the Viking landers on Mars. Very precise orbital data, beginning in 1974, also exist for the binary pulsar PSR 1913+16. A weaker but also less model-dependent bound can be derived from the spin-down rate of the pulsars JP 1953 (Heintzmann & Hillebrandt 1975) and PSR 0655+64 (Goldmann 1990). The resulting present-day constraints on $\dot G/G$ are summarized in Table 1. More detailed discussions can be found in the book by Will (1993).

**Table 1.** Celestial-mechanics bounds on the present-day $\dot G/G$. (Adapted from Will 1993.)

| Method | $\dot G/G$ [$10^{-12}$ yr$^{-1}$] | References |
|---|---|---|
| Laser ranging (Moon) | $0 \pm 10$ | Müller et al. (1991) |
| Radar ranging (Mars) | $-2 \pm 10$ | Shapiro (1990) |
| Binary pulsar 1913+16 | $11 \pm 11$ | Damour & Taylor (1991) |
| Spin-down PSR 0655+64 | $< 55$ | Goldman (1990) |

### 2.2. Big Bang Nucleosynthesis (BBN)

A constraint on the value of $G$ in the early universe arises from the observed primordial light element abundances (Barrow 1978).[1] In a Friedman-Robertson-Walker model of the universe the expansion rate is given by $H^2 = \frac{8\pi}{3} G \rho$ in terms of the energy density $\rho$ which, during the epoch of nucleosynthesis, is dominated by radiation (photons, neutrinos). It is a standard argument to constrain $\rho$ from the yield of $^4$He and other light elements, and thus to constrain the effective number of neutrino degrees of freedom at nucleosynthesis (Yang et al. 1979, 1984; Olive et al. 1990; Walker et al. 1991). Because the number of low-mass sequential neutrino families is now known to be 3, $\rho$ on the r.h.s. of the Friedman equation is fixed. Therefore, barring novel particle-physics effects which could still modify $\rho$, one may constrain the value of $G$ at the BBN epoch.

Actually, the consistency of the BBN predictions of the light element abundances with observations is a topic of current debate (Olive & Scully 1995). Hata et al. find that the observationally inferred deuterium and helium abundances would be significantly more consistent with each other if there were only 2 neutrino flavors. Because we know that there are 3 flavors, and because one neutrino species contributes around 15% to $\rho$, a reduced $G$ value by something like 15% would be favored.

Nevertheless, BBN probably excludes an $\mathcal{O}(1)$ deviation of $G$; for definiteness we assume that $G$ was within $\pm 50\%$ of its present value. This constraint can be compared with the above celestial-mechanics bounds only by assuming a specific functional form for $G(t)$. We will usually take a linear dependence

$$G(t) = G_0 \left[ 1 + \Gamma_0 (t - t_0) \right] \quad (1)$$

where $t = t_0$ is the cosmic time at the present epoch, $G_0$ is the present-day value of Newton's constant, and $\Gamma(t) \equiv \dot G(t)/G_0$. For a linear $G(t)$ variation, of course, $\Gamma(t) = \Gamma_0$ is constant. BBN then implies $|\dot G_0/G_0| = |\Gamma_0| \lesssim 30 \times 10^{-12}$ yr$^{-1}$, somewhat less restrictive than the celestial-mechanics limits.

Several authors assumed that $G$ varies according to a power law

$$G(t) = G_0 (t/t_0)^\beta \quad (2)$$

where $t = 0$ refers to the big bang. In this case the slope is $\Gamma(t) \equiv \dot G(t)/G_0 = (\beta/t_0)(t/t_0)^{\beta-1}$ so that today $\Gamma_0 = \beta/t_0$. BBN yields $|\beta| \lesssim 0.01$ or $\Gamma_0 \lesssim 10^{-12}$ yr$^{-1}$, at least a factor of ten below the celestial-mechanics limits.

While a power-law or a linear variation of $G$ are both relatively arbitrary assumptions, it is still noteworthy that the sensitivities of the BBN and the celestial-mechanics methods to $\dot G$ are of the same general order of magnitude. Both approaches leave room for a considerable $G$ variation over cosmic time scales.

### 2.3. Properties of the Sun

If $G$ did vary in time one would expect a modification of the standard course of stellar evolution (Teller 1948). By means of a beautifully simple homology argument Teller showed that the luminosity of the Sun is approximately proportional to $G^7$ (see also Appendix A). The existence of life on Earth during the past 500 million years or more then allowed Teller to constrain a possible $G$ variation. On the basis of a homology argument and using the solar age as an indicator Gamow (1967) derived related limits. In the sixties when the Brans-Dicke theory was in vogue, detailed models of the Sun with a varying $G$ were constructed by Pochoda & Schwarzschild (1964), Ezer & Cameron (1966), Roeder & Demarque (1966), Shaviv & Bahcall (1969), and later by Chin & Stothers (1975, 1976).

The crux with constraining $\dot G$ from the Sun is that the presolar helium abundance and the mixing length parameter can and must be tuned to reproduce the Sun's present-day luminosity and radius. Even though the present-day central temperature, density, and helium abundance could differ vastly from standard predictions, their main impact would be on the neutrino

---
[1] Apparently there is an earlier discussion of this limit by G. Steigman in an unpublished essay for the 1976 Gravity Research Foundation Awards. Subsequent refinements include Rothman & Matzner (1982), Accetta, Krauss & Romanelli (1990), Damour & Gundlach (1991), and Casas, García-Bellido & Quirós (1992).

conditions as it may get modified by neutrino oscillations.

At the present time the most sensitive probe of a variant internal solar structure is afforded by the measured p-mode frequencies which agree well with standard predictions, especially when the gravitational settling of helium is taken into account. Recently Demarque et al. (1994) have constructed solar models with a varying $G$ and then analyzed their p-mode spectra in comparison with the observations. They conclude that

$$|\dot{G}/G|_0 \lesssim 30\times 10^{-12}\,\text{yr}^{-1} \qquad (3)$$

is a reasonably conservative limit. It is similar to the celestial-mechanics bounds of Table 1.

A similar study was recently completed by Guenther et al. (1995) who focussed on the predicted g-mode spectrum. At the present time there is no generally accepted observation of solar g-modes. If one were to take the claimed observations by Hill & Gu (1990) seriously, a bound $|\dot{G}/G|_0 \lesssim 3\times 10^{-12}\,\text{yr}^{-1}$ would ensue. Therefore, if an unambiguous identification of g-modes would emerge from a number of forthcoming observational projects, the Sun may yet provide one of the most restrictive limits on the constancy of Newton's constant.

### 2.4. White dwarfs

The largest impact of a time-varying gravitational constant can be expected on the oldest stellar objects which "integrate" $G(t)$ into the more distant past than does the evolution of the Sun. One well understood category of such objects are white dwarfs, the faintest of which likely formed shortly after the birth of the galactic disk. Therefore, the age of the galactic disk implied by the fast drop of the white dwarf luminosity function at the faint end depends on the $G$ evolution in the past.

An early study of white dwarf properties if $G$ varies in time was performed by Vila (1976) who concluded on the basis of the observations available at that time that a rate of change as large as $75\times 10^{-12}\,\text{yr}^{-1}$ was not excluded. In a recent investigation by García-Berro et al. (1995), detailed luminosity functions were constructed under the assumption of a decreasing $G$. For an assumed age of the galactic disk of 7 Gyr, which probably is a lower plausible limit, the best fit for the faintest data point requires $\dot{G}/G = -10\times 10^{-12}\,\text{yr}^{-1}$ while the curves for 0 and $-30\times 10^{-12}\,\text{yr}^{-1}$ lie somewhat outside of the $1\sigma$ error bar of this all-important data point. Therefore, the white dwarf luminosity function at present does not seem to yield significant limits on the scale of the above celestial-mechanics results.

### 2.5. Globular clusters

One may be able to do better with globular cluster (GC) stars. A color-magnitude diagram for an intermediate-aged galactic cluster was constructed by Roeder (1967) while a detailed study of GCs was performed by Prather (1976). In both works a time variation for a specific Brans-Dicke cosmology was assumed where $G$ decreases approximately as in Eq. (2) with $\beta \approx 0.03$. Prather concluded that it was difficult to find evidence for a to extract limits on a generic $G$ time variation from his results. Therefore, we presently reexamine the impact of a time-varying gravitational constant on GCs.

## 3. Time-varying gravity and globular cluster stars

### 3.1. Color-magnitude diagram

In order to gain some intuition for the evolution of low-mass stars which experience a time-varying gravitational constant we have evolved numerically several stars from the zero age main sequence to the red giant branch. The evolution after the subgiant phase is fast on cosmological time scales so that we expect, in accordance with Prather's (1976) findings, that whatever the impact of a time-varying $G$ on stellar evolution, the advanced stages of low-mass stars will not be affected by the past history of $G(t)$. Therefore, it is not necessary to calculate up to the helium flash or beyond.

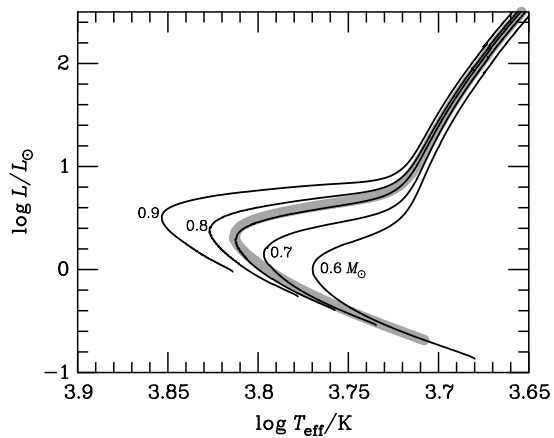

**Fig. 1.** Reference evolutionary tracks, labelled with the stellar mass. The 0.75 $M_\odot$ track almost coincides with the 16 Gyr isochrone (thick shaded line)

The numerical code we use (FRANEC) and its main physical inputs have been described elsewhere (e.g. Chieffi & Straniero 1989; Castellani, Chieffi & Pulone 1991). We used the OPAL opacities (e.g. Rogers & Iglesias 1992) for the stellar interior, and the Alexander & Ferguson (1994) opacities for low-temperature regions.

For the stellar parameters we have chosen a metallicity $Z = 2\times 10^{-4}$ and an initial helium abundance $Y_\text{init} = 0.24$ or $= 0.25$, which are typical for halo GCs in our galaxy. As a measured value for the bolometric luminosity of the main-sequence (MS) turnoff $L_\text{TO}$ in a typical GC we use $\log(L_\text{TO}/L_\odot) = 0.30$ with $L_\odot$ the solar luminosity and $\log \equiv \log_{10}$. For these parameters the turnoff mass is found to be $M_\text{TO} = 0.75\,M_\odot$. In the constant-gravity framework this corresponds to a GC age of 16 Gyr. In Fig. 1 we illustrate this reference case with several evolutionary tracks as well as the 16 Gyr isochrone (thick shaded line).

cording to Eq. (1). In Fig. 2 we present the evolutionary tracks of a 0.75 $M_\odot$ star for constant standard gravity (STD) as in Fig. 1, and further a track (a) with decreasing, and one (b) with increasing gravity. For (a) we took $\Gamma_0\tau = -0.10$ where $\tau$ is the time it took the star to reach the turnoff. Put another way, we had to adjust $\Gamma_0$ such that the 0.75 $M_\odot$ star would reach its MS turnoff at the present epoch which is defined by $G(t) = G_0$. We found $\tau = 12$ Gyr and thus $\Gamma_0 = -8.3\times 10^{-12}$ yr$^{-1}$. For (b) we took $\Gamma_0\tau = +0.08$ which led to $\tau = 20$ Gyr and $\Gamma_0 = +4\times 10^{-12}$ yr$^{-1}$. The interpretation of this is that with a $\Gamma_0 = -8.3\times 10^{-12}$ yr$^{-1}$ decrease of gravity a star of only 12 Gyr true age would reach the same turnoff position as one of 16 Gyr evolving at constant $G$.

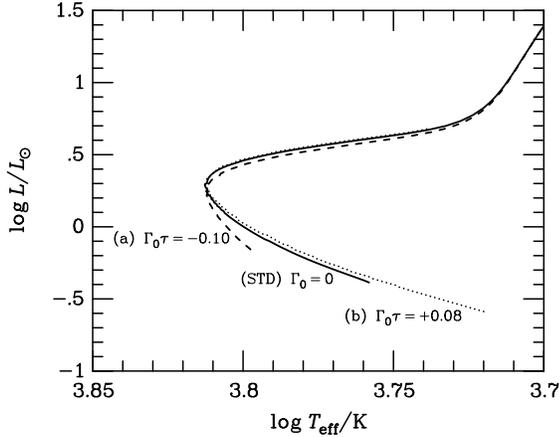

**Fig. 2.** $M = 0.75\,M_\odot$ evolutionary tracks. *(STD)* Standard constant gravity case; the MS turnoff occurs at 16 Gyr. *(a)* Linearly decreasing gravity with $\Gamma_0\tau = -0.10$. *(b)* Linearly increasing gravity with $\Gamma_0\tau = +0.08$.

In these cases the TO brightness is unchanged to better than 0.1 mag which corresponds to the observational uncertainty. We conclude that, at least for moderate variations of $G$ (as in Table 2 below), the stars at TO retain little memory of the past $G(t)$ history, i.e. their properties are determined almost exclusively by the present-day strength of gravity.

We stress, however, that the evolutionary tracks are significantly distorted if the $G$ variation is stronger by a factor of 2–3 (see also Roeder & Demarque 1966; Roeder 1967). This is illustrated in Fig. 3 for a linear $G(t)$ variation with $\Gamma_0 = -17.8\times 10^{-12}$ yr$^{-1}$. We show the evolutionary tracks for the indicated stellar masses and with $Z = 2\times 10^{-4}$ and $Y_{\rm init} = 0.25$. (For these parameters the observed TO luminosity corresponds to a 0.76 $M_\odot$ star which for constant gravity turns off the MS at 14 Gyr.) The stars of Fig. 3 were born 9 Gyr before gravity reached its present-day value. Therefore, at the birth of these stars it was 16% stronger than it is today. This true cluster age was chosen such that the 0.76 $M_\odot$ stars turn off the MS at the present epoch as they would for constant gravity. Beyond the present epoch the tracks describe the future evolution (dot-with the lowest masses will never ascend the giant branch.

The thick shaded line is the standard constant-gravity isochrone for this cluster. Evidently, it is nearly identical with the present-day isochrone of the varying-gravity cluster (bullets in Fig. 3, i.e. the 9 Gyr location on the tracks). Actually, for standard ages greater than a few Gyr the nonstandard isochrones already look very similar to the standard ones. Even in the extreme example of Fig. 3 there is no obvious signature in the GC color-magnitude diagram for the past variation of the gravitational constant.

In summary, we confirm the finding of previous authors that the properties of the GC isochrones are not dramatically changed by a time variation of Newton's constant. Apparently, then, the modified evolutionary time scale on the MS is the only significant consequence. This implies that as long as no other restriction on the true cluster age is available (e.g. by cosmological arguments concerning the age of the universe as a whole or by additional and independent age indicators for GCs) GC isochrones cannot be used to restrict $\dot{G}/G$.

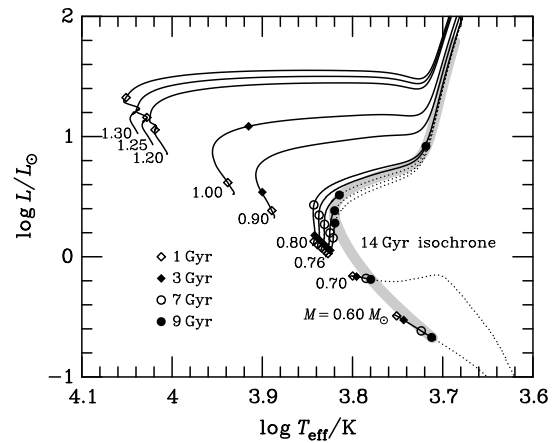

**Fig. 3.** Evolutionary tracks for linearly decreasing gravity with $\Gamma_0 = -17.8\times 10^{-12}$ yr$^{-1}$. The true cluster age was chosen as 9 Gyr so that a 0.76 $M_\odot$ star ($Y_{\rm init} = 0.25$) turns off the MS today with the observed TO luminosity. For constant gravity this would occur at 14 Gyr. The future evolution when gravity will become weaker is shown as dotted lines. The masses of the models and several locations of constant age are indicated. The thick shaded line is the standard isochrone of the constant-gravity cluster which correponds well to the present-day varying-gravity isochrone indicated by the bullets for 9 Gyr.

### 3.2. Main-sequence evolutionary time scale

#### 3.2.1. General expression

If gravity varies in time the modification of the MS evolutionary time-scale can be estimated by a surprisingly accurate analytical method. To this end we note that the MS luminosity of a stellar model of given mass and initial composition is fixed by the magnitude of Newton's constant $G$ and by the age of the drogen to helium in its core. If $Y$ denotes the helium abundance in the hydrogen-burning central regions one may assume that approximately $L \propto f(Y)h(G)$ where $f$ and $h$ are some functions of $Y$ and $G$, respectively. The assumption that $L$ depends only on the instantaneous value of $G$ amounts to the assumption that the star at no time retains a significant memory of its past history, except by the amount of hydrogen burnt. This assumption is justified by the adiabatic $G(t)$ variation which varies on cosmological time scales which are slower than any time scale that governs stellar evolution. Because helium is produced at a rate proportional to $L$ we have $dY/dt \propto f(Y)h(G)$ or $dY/f(Y) \propto dt\, h[G(t)]$. A star which today ($t = t_0$) turns off the MS has at its center $Y \approx 1$ so that

$$\int_{Y_{\rm init}}^{1} \frac{dY}{f(Y)} \propto \int_{t_{\rm init}}^{t_0} dt\, h[G(t)] \qquad (4)$$

where $t_{\rm init}$ is the time the star was born. The l.h.s. does not depend on $G$ whence the r.h.s. is the same for any function $G(t)$.

As a further approximation we assume that $h(G)$ is a power law $G^\gamma$. The power $\gamma$ is found to be about 5.6 by a homology argument together with a numerical calibration for our chemical composition of interest (Appendix A). It follows that

$$\tau_* = \int_{t_0-\tau}^{t_0} dt\, [G(t)/G_0]^\gamma . \qquad (5)$$

Here $\tau_*$ is the apparent turnoff age of a star, i.e. the age it would have in a constant-gravity scenario, while $\tau$ is its true age if gravity varies according to $G(t)$. An equivalent equation was previously stated by Prather (1976).

### 3.2.2. Linear $G(t)$ variation

If we take explicitly the linear $G(t)$ dependence of Eq. (1) we find with $p \equiv \gamma + 1$

$$\Gamma_0 \tau_* = \frac{1 - (1 - \Gamma_0 \tau)^p}{p} \qquad (6)$$

or equivalently

$$\frac{\tau}{\tau_*} = \frac{p\, \Gamma_0 \tau}{1 - (1 - \Gamma_0 \tau)^p},$$
$$\frac{\tau}{\tau_*} = \frac{1 - (1 - p\,\Gamma_0 \tau_*)^{1/p}}{\Gamma_0 \tau_*}. \qquad (7)$$

For $\gamma = 5.6$ (Appendix A), i.e. $p = 6.6$, we show $\Gamma_0 \tau_*$ (solid line) in Fig. 4 as a function of $\tau/\tau_*$. To linear order in $\tau/\tau_*$ both results expand as $\Gamma_0 \tau_* = \Gamma_0 \tau = (2/\gamma)(\tau/\tau_* - 1)$; see the dotted line in Fig. 4.

In order to check the precision of these analytical results we considered numerically the effect of a linear $G(t)$ variation on our $0.75\,M_\odot$ star which has $\tau_* = 16\,{\rm Gyr}$. We determined $\Gamma_0$ such that the true age at the MS turnoff would be 12 and 20 Gyr, respectively (Table 2). There is good agreement with the analytical formula—see the squares in Fig. 4.

**Table 2.** Numerical calculations with linearly varying gravity as described in the text.

| $M$ [$M_\odot$] | $Y_{\rm init}$ | $\tau_*$ [Gyr] | $\tau$ [Gyr] | $\Gamma_0$ [$10^{-12}$ yr$^{-1}$] | $\log(L_{\rm TO}/L_\odot)$ |
|---|---|---|---|---|---|
| 0.75 | 0.24 | 16 | 12 | $-8.3$ | 0.26 |
| 0.75 | 0.24 | 16 | 20 | $+4$ | 0.29 |
| 0.76 | 0.25 | 14 | 12 | $-4.2$ | 0.33 |
| 0.76 | 0.25 | 14 | 10 | $-11$ | 0.29 |

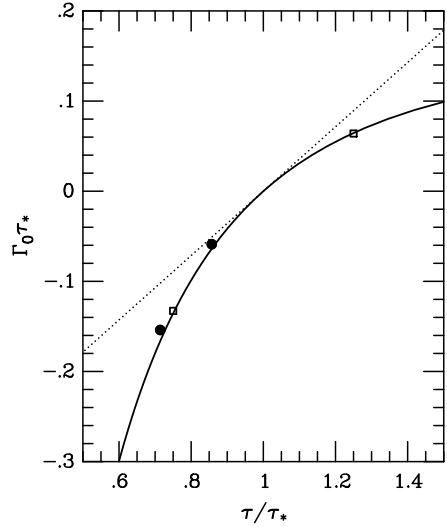

**Fig. 4.** $\Gamma_0 \tau_*$ as a function of $\tau/\tau_*$ according to Eq. (7). The dotted line corresponds to the linear approximation at $\tau/\tau_* = 1$. The squares correspond to the numerical results of Table 2 with $\tau_* = 16\,{\rm Gyr}$, the bullets to those with $\tau_* = 14\,{\rm Gyr}$.

As a last case we constructed a model which in the constant-gravity case yields the smallest apparent GC age compatible with the uncertainties in the observational determination of $L_{\rm TO}$ and $Y_{\rm init}$. To this end we held the metallicity fixed at $Z = 2\times 10^{-4}$, took $Y_{\rm init} = 0.25$, and $L_{\rm TO}$ 10% higher than in previous calculations which is consistent with the observational uncertainty. We then find an apparent GC age of 14 Gyr and a turnoff mass of $0.76\,M_\odot$ (instead of 16 Gyr and $0.75\,M_\odot$). This confirms the well known result that the GC age can be shortened by about 2 Gyr by changing $Y_{\rm init}$ and $L_{\rm TO}$ within the estimated errors.

Once more we checked the above analytical results by determining the values for $\Gamma_0$ which are required to change $\tau_* = 14\,{\rm Gyr}$ to $\tau = 10$ and 12 Gyr, respectively (Table 2). Again, there is excellent agreement with the analytical formula (bullets in Fig. 4).

One may go through the same exercises for the power-law $G(t)$ variation of Eq. (2) which leads to

$$\frac{\tau}{\tau_*} = \frac{(1+\beta\gamma)\,\tau/t_0}{1-(1-\tau/t_0)^{1+\beta\gamma}}. \tag{8}$$

GCs were formed only a few Gyr after the big bang, such that one has $\tau/t_0 \approx 1$. With $\Gamma_0 = \beta/t_0$ one finds in this limiting case explicitly

$$\Gamma_0 = \frac{1}{\gamma}\left(\frac{1}{\tau_*} - \frac{1}{\tau}\right) \tag{9}$$

for the required present-day $\dot G/G$ to achieve a desired GC age $\tau$. This expression is a reasonable approximation even if GCs were born several Gyr after the big bang.

We have tested the accuracy of the analytical result Eq. (8) numerically with the $0.76\,M_\odot$ star which has $\tau_* = 14$ Gyr. With $\tau/t_0 = 0.90$ we found the $\beta$ values listed in Table 3 in order to have $t_0 = 12$ and $20$ Gyr, respectively. Again, the analytic formula works exceedingly well.

**Table 3.** Numerical calculations with a power-law variation of $G(t)$ as described in the text.

| $M$ [$M_\odot$] | $Y_{init}$ | $\tau_*$ [Gyr] | $\tau$ [Gyr] | $t_0$ [Gyr] | $\beta$ | $\log(L_{TO}/L_\odot)$ |
|---|---|---|---|---|---|---|
| 0.76 | 0.25 | 14 | 18 | 20 | +0.07 | 0.32 |
| 0.76 | 0.25 | 14 | 10.8 | 12 | −0.06 | 0.32 |

Therefore, for all practical purposes the analytical approach to estimating the modification of the GC ages due to a time-varying gravitational constant works extremely well.

### 3.3. Globular-cluster bound on present-day $\dot G/G$

We are now in a position to use these results to derive a new limit on the present-day $\dot G/G$ from GC ages. To this end we plot in Fig. 5 the required present-day $\dot G/G$ as a function of the desired age of GCs. The solid lines are for a linear $G(t)$ variation, taking $\tau_* = 14$ and $18$ Gyr for the apparent GC age, respectively. The dotted lines are for a power-law variation, the same apparent ages, and $\tau/t_0 = 1$ so that the GCs were born immediately after the big bang. Allowing for a time delay of a few Gyr does not change these curves very much. They are always between the corresponding dotted and solid lines in agreement with the intuition that the power law with a delayed GC birth approaches the linear-variation case. In this sense the power law with $\tau/t_0 = 1$ and the linear case represent the two extremes of GCs being born immediately after and very long after the big bang, respectively.

From Fig. 5 it is evident that for a given range of allowed GC ages the linear time variation yields less restrictive limits on the present-day variation of Newton's constant. The expansion age of the universe is generously limited to be less than 20 Gyr, and because globular clusters cannot be any older one finds that today

$$\dot G/G \lesssim 7\times 10^{-12}\,\mathrm{yr}^{-1}, \tag{10}$$

i.e. we find a very restrictive limit on the possibility that gravity has been weaker in the past. (Of course, the expansion age of the universe would also be affected by a time-varying gravitational constant. However, because the expansion parameter scales as $G^{1/2}$ while the stellar evolutionary time scale as $G^{5.6}$ we neglect this effect.) Conversely, a moderately stronger $G$ in the past would suffice to lower GC ages below, say, 12 Gyr. However, our method does not allow us to derive a significant limit on the possibility that $G$ was indeed stronger. To this end one would need a significant lower limit to GC ages. If we were sure that $\tau \gtrsim 8$ Gyr (e.g. because $H_0 \lesssim 80\,\mathrm{km\,s^{-1}\,Mpc^{-1}}$ in a closed univerise) we would have $\dot G/G \gtrsim -35\times 10^{-12}\,\mathrm{yr}^{-1}$ which is not more restrictive than what is known, say, from the solar p-mode frequencies.

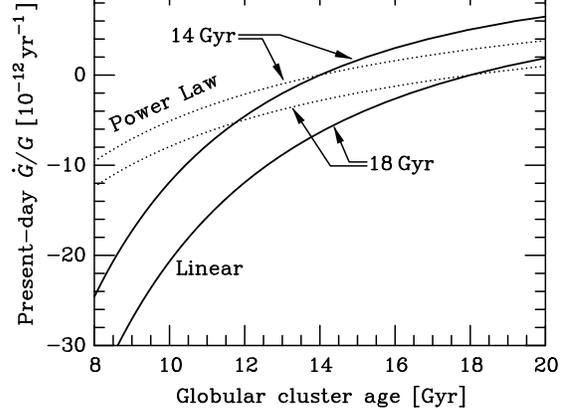

**Fig. 5.** Required present-day $\dot G/G$ in order to achieve a desired GC age. For both the linear $G(t)$ variation (solid lines) and the power law (dotted lines) two apparent ages 14 and 18 Gyr were used. For the power-law the age of the universe $t_0$ was taken to be equal to the true GC age $\tau$.

### 4. Summary

In agreement with Prather's (1976) previous findings we have confirmed that the dominant impact of a time-varying gravitational constant on GCs is a modification of the MS evolutionary time scale while the appearance of the color-magnitude diagram remains undistorted within the observational resolution and within theoretical uncertainties. The evolutionary MS time scale can be calculated by the very precise analytic approximation Eq. (5). Since GCs must be younger than the universe as a whole, age determinations independent of GCs —e.g. by determination of the Hubble constant— can thus be used to limit the variation of $G$. If GCs are younger than 20 Gyr, which is the determination, we find that the present-day rate of change $\dot{G}/G$ must be less than $7\times10^{-12}$ yr$^{-1}$, i.e. gravity cannot have been much weaker in the past. This conclusion is based on a linear $G(t)$ variation. An assumed power-law variation yields an even more restrictive limit. If gravity was slightly stronger in the past, GCs can be made younger than they look, allowing one to cure the discrepancy between the apparent GC ages and the cosmic expansion age, which might be close to 10 Gyr according to recent $H_0$ determinations (Freedman et al. 1994; Mould et al. 1995). We are not able to derive a significant limit on $\dot{G}/G$ if gravity was indeed stronger in the past, but we demonstrated that a linear rate of change of $\dot{G}/G = -8\times10^{-12}$ yr$^{-1}$ would make a GC of 12 Gyr (true age) appear as a standard 16 Gyr one.

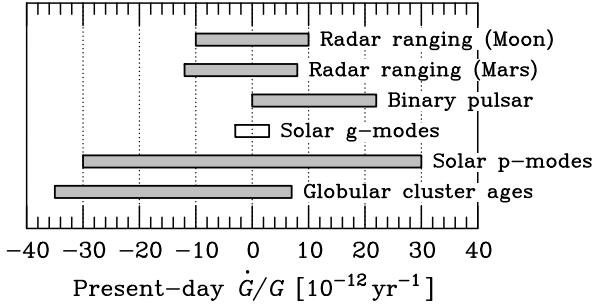

**Fig. 6.** Limits on the present-day $\dot{G}/G$ from celestial mechanics (Table 1), from helioseismology (Sect. 2.3), and our new globular cluster limit.

In Fig. 6 we summarize the available limits on a present-day $G$ time variation. We do not show the BBN limit because its meaning for the present-day value of $\dot{G}/G$ depends almost entirely on the assumed $G(t)$ behavior at early times. For $\dot{G}/G > 0$ (gravity weaker in the past) our limit is competitive with those from celestial mechanics of Table 1, and significantly better than those from the solar p-mode frequencies. However, even if a definitive discovery of solar g-modes would confirm the shown tentative limit one could not exclude that GCs look older than they are because in the power-law $G(t)$ scenario a rather small present-day $\dot{G}/G$ would ensue. Current $\dot{G}/G$ limits would have to be improved by, say, an order of magnitude to eliminate a time variation of Newton's constant as an explanation for the old looks of globular clusters.

*Acknowledgements.* This research was partially supported by the European Union contract CHRX-CT93-0120 and by the Deutsche Forschungsgemeinschaft grant SFB-375. G.F. and B.R. enjoyed the hospitality of the MPI für Astrophysik while this work was conducted.

## A. Homology relations

The main sequence (MS) luminosity as a function of stellar mass and gravitational constant can be estimated by using homology sider the core of a metal poor MS star of about $1\,M_\odot$ where the central density is about $100$ g cm$^{-3}$ and the central temperature about $10^7$ K. For such conditions we simplify the equation of state by the ideal-gas law $P \propto \rho T$, while the energy-generation rate $\epsilon$ is dominated by the pp chains whence $\epsilon \propto \rho T^4$. Energy is transported by radiation and we assume that the opacity per unit mass can be expressed as

$$\kappa \propto \rho^\delta/T^\nu \tag{A1}$$

where $\delta$ and $\nu$ are dimensionless parameters.

Together with the four stellar equilibrium conditions, the five scales for luminosity, radius, temperature, pressure, and density ($L$, $R$, $T$, $P$, $\rho$) can all be expressed in terms of the gravitational constant $G$ and the stellar mass $M$. In particular, by writing

$$L \propto G^\gamma M^\mu \tag{A2}$$

one finds

$$\gamma = \frac{28 + 12\delta + 3\nu}{7 + 3\delta - \nu}$$
$$\mu = \frac{21 + 11\delta + \nu}{7 + 3\delta - \nu} \tag{A3}$$

The presence of $\delta$ and $\nu$ in these relations reveals the importance of the opacity.

For our conditions of interest the opacity is essentially given by Thomson scattering and free-free transitions so that $\kappa = \kappa_{\rm Th} + \kappa_{\rm ff} = c_1 + c_2 \rho T^{-3.5}$. For $\rho = 100$ g cm$^{-3}$ and $T = 10^7$ K analytical estimates give (e.g. Kippenhahn & Weigert 1990) $\kappa_{\rm Th} = 0.2\,(2-Y)$ and $\kappa_{\rm ff} = 1.2\,(2-Y)$ where cgs units are understood. Therefore, $\kappa_{\rm Th}$ and $\kappa_{\rm ff}$ are comparable in our region of interest. With Eq. (A1) one finds

$$\delta = \frac{\kappa_{\rm ff}}{\kappa_{\rm ff} + \kappa_{\rm Th}},$$
$$\nu = 3.5\,\delta. \tag{A4}$$

Therefore, $0 < \delta < 1$ and $\nu < 3.5$, the precise values depending on $\kappa_{\rm Th}/\kappa_{\rm ff}$.

For pure Thomson scattering one has $\delta = \nu = 0$ and so $L \propto G^4 M^3$. Pure free-free transitions ($\delta = 1$, $\nu = 3.5$) yield $L \propto G^{7.8} M^{5.5}$; this was Gamow's (1967) choice. Finally, Teller (1948) took $\delta = 1$ and $\nu = 3$ which yields $L \propto G^7 M^5$. In our evolutionary code we use the Livermore opacity tables (OPAL). Close to the center of a $0.76\,M_\odot$ metal poor zero-age MS star with Y=0.25 they can be approximated by $\delta = 0.5$ and $\nu = 1.75$. For the same star we find at the MS turnoff $\delta = 0.55$ and $\nu = 2.2$. For studying the MS evolution we consider the zero-age values to be more representative. They yield $L \propto G^{5.8} M^{4.2}$. Within 10% this agrees with our direct numerical results

$$L \propto G^{5.6} M^{4.7} \tag{A5}$$

which we have obtained with the Frascati evolution code (FRANEC).